\documentclass[11pt]{article}
\usepackage{latexsym,amssymb} 
\newcommand{\ft}[2]{{\textstyle\frac{#1}{#2}}}

\def\bfone{\relax{\rm 1\kern-.35em 1}}

\newcommand{\llceil}{{|\!\!|\!\!\lceil}}
\newcommand{\rrfloor}{{\rfloor\!\!|\!\!|}}

\newdimen\squaresize \squaresize=12pt
\newdimen\thickness \thickness=0.7pt

\def\square#1{\hbox{\vrule width \thickness
   \vbox to \squaresize{\hrule height \thickness\vss
      \hbox to \squaresize{\hss#1\hss}
   \vss\hrule height\thickness}
\unskip\vrule width \thickness} \kern-\thickness}

\def\cut#1{\hbox{\vrule width-1 \thickness
   \vbox to \squaresize{\hrule height-1 \thickness\vss
      \hbox to \squaresize{\hss#1\hss}
   \vss\hrule height-1\thickness}
\unskip\vrule width +4 \thickness} \kern-\thickness}

\def\vsquare#1{\vbox{\square{$#1$}}\kern-\thickness}

\def\young#1{
\vbox{\smallskip\offinterlineskip \halign{&\vsquare{##}\cr #1}}}

\newcommand{\tinyyoung}[1]{
\squaresize=7pt \thickness=0.4pt \mbox{\tiny\young{#1}}
\squaresize=12pt \thickness=0.7pt}

\begin{document}
\begin{titlepage}

\begin{flushright}\small 
LPTENS-08/26\\ITP-UU-08/30\\ SPIN-08/23\\ENSL-00283596
\end{flushright}
%
\vskip 10mm
\begin{center}
  {\huge {\bf The end of the {\em p}\,-form hierarchy}}
\end{center}
\vskip 8mm

\begin{center}
{\Large{\bf Bernard de Wit${}^{\rm a,b}$ and Henning Samtleben${}^{\rm c}$}}
  \\[6mm]

${}^a$ Laboratoire de Physique Th\'eorique\\ de l'Ecole Normale
  Sup\'erieure, CNRS-UMR8549, \\
  24 rue Lhomond, F-75231 Paris Cedex 05, France\\[2mm]
${}^b$ Institute for Theoretical Physics \,\&\, Spinoza Institute,\\  
Utrecht University, P.O. Box  80.195, NL-3508 TD Utrecht, 
The Netherlands\\[2mm]
${}^c$ Universit\'e de Lyon, Laboratoire de Physique, \\
Ecole Normale Sup\'erieure de Lyon,\\ 
46 all\'ee d'Italie, F-69364 Lyon Cedex 07, France \\[3mm]
{${\tt b.dewit@uu.nl}\;,\;{\tt henning.samtleben@ens-lyon.fr}$}
\end{center}

\vskip .2in

\begin{center} {\bf Abstract } 
\end{center}
\begin{quotation}\noindent
  The introduction of a non-abelian gauge group embedded into the
  rigid symmetry group $\mathrm{G}$ of a field theory with abelian
  vector fields and no corresponding charges, requires in general the
  presence of a hierarchy of $p$-form gauge fields. The full gauge
  algebra of this hierarchy can be defined independently of a specific
  theory and is encoded in the embedding tensor that determines the
  gauge group. When applied to specific Lagrangians, the algebra is
  deformed in an intricate way and in general will only close up to
  equations of motion. The group-theoretical structure of the
  hierarchy exhibits many interesting features, which have 
  been studied starting from the low-$p$ forms. Here the question is
  addressed what happens generically for high values of $p$. In
  addition a number of other features is discussed concerning the role
  that the $p$-forms play in various deformations of the theory.
\end{quotation}
\end{titlepage}
\eject  
\section{Introduction}
\setcounter{equation}{0}
\label{sec:introduction}

In recent years the study of general gaugings of extended
supergravities initiated in \cite{deWit:2004nw,deWit:2005hv} has led
to considerable insight in the general question of embedding a
non-abelian gauge group into the rigid symmetry group $\mathrm{G}$ of
a theory that contains abelian vector fields without corresponding
charges, transforming in some representation of $\mathrm{G}$ (usually
not in the adjoint representation). The field content of this theory
is fixed up to possible dualities between \mbox{$p$-forms} and
$(d-p-2)$-forms. Therefore, it is advantageous to adopt a framework in
which the decomposition of the form fields is determined only until
after the gauging. The relevance of this can, for instance, be seen in
four space-time dimensions \cite{deWit:2005ub}, where the Lagrangian
can be changed by electric/magnetic duality so that electric gauge
fields are replaced by their magnetic duals. In the usual setting, one
has to adopt an electric/magnetic duality frame where the gauge fields
associated with the desired gauging are all electric. In principle
this may not be sufficient, because the gauge fields should decompose
under the embedded gauge group into fields transforming in the adjoint
representation of the gauge group, and fields that are invariant under
this group, so as to avoid inconsistencies.  In a more covariant
framework, on the other hand, one introduces both electric and
magnetic gauge fields from the start, such that the desired gauge
group can be embedded irrespectively of the particular electric/magnetic
duality frame. Gauge charges can then be switched on in a fully
covariant setting provided one introduces 2-form fields transforming
in the adjoint representation of $\mathrm{G}$. To keep the number of
physical degrees of freedom unchanged, new gauge transformations
associated with the 2-form gauge fields are necessary.  In this
approach the gauge group embedding is encoded in the so-called
embedding tensor, which is treated as a spurionic quantity so as to
make it amenable to group-theoretical methods.

This group-theoretical framework has already been applied to a rather
large variety of supergravity theories in various space-time
dimensions, where it was possible to characterize all possible gauge
group embeddings in a group-theoretical fashion
\cite{Samtleben:2005bp,Schon:2006kz,deWit:2007mt,
  deVroome:2007zd,Derendinger:2007xp,Samtleben:2007an,Bergshoeff:2007ef}.
We note that the three-dimensional theories
\cite{Nicolai:2000sc,Nicolai:2001sv,deWit:2004yr} also fall in this
class, although they are special in that the vector gauge fields
themselves can be avoided in the absence of any gauging, as they are
dual to scalar fields. It is in this context that the embedding tensor
was first introduced.

While in four space-time dimensions no $p$-form fields are required in
the action beyond ${p=2}$, the higher-dimensional case may incorporate
higher-rank form fields which will naturally extend to a hierarchy when
switching on gauge charges, inducing a non-trivial entanglement
between forms of different ranks.  It may seem that one introduces an
infinite number of degrees of freedom in this way, but, as mentioned
already above, the hierarchy contains additional gauge invariances
beyond those associated with the vector fields.  As it turns out, this
hierarchy is entirely determined by the rigid symmetry group $\mathrm{G}$
and the embedding tensor that defines the gauge group embedding into
$\mathrm{G}$ \cite{deWit:2005hv,deWit:2008ta} and a priori makes
neither reference to an action nor to the number $d$ of space-time
dimensions.  In particular, as a group-theoretical construct, the
tensor hierarchy in principle continues indefinitely, but it can be
consistently truncated in agreement with the space-time properties
(notably the absence of forms of a rank $p>d$). In this paper we will
analyze some of the generic features of the hierarchy for large values
of the rank $p$ (i.e.\ close to $d$).

Although every choice of embedding tensor defines a particular gauging
and thereby a corresponding $p$-form hierarchy, it turns out that the
hierarchy is universal in the sense that scanning through all possible
choices of the embedding tensor and taking into account the
group-theoretical representation constraints which it obeys, allows to
characterize the multiplicity of the various $p$-forms in entire
$\mathrm{G}$-representations -- within which every specific gauging
selects its proper subset. This is precisely the meaning of treating
the embedding tensor as a spurionic quantity. The covariant
description of the gauged tensor hierarchy thereby enables the
derivation of the full $p$-form field content. In the general case it
may be difficult to indicate the precise $\mathrm{G}$-representations
to which the $p$-forms are assigned, but it is possible to indicate
all the ingredients in a systematic way (although it requires some
notational ingenuity) such that they can be worked out explicitly on a
case-by-case basis~\cite{deWit:2008ta}.

Although the structure of the $p$-form hierarchy seems to be
universal, the situation changes when incorporating this formalism in
the context of a given Lagrangian. The transformation rules are then
deformed by the presence of the various matter fields and, as a
result, the closure of the generalized gauge algebra only holds up to
equations of motion and additional symmetries (which are connected to
certain redundancies in the transformation rules of the
hierarchy)~\cite{deWit:2007mt,deVroome:2007zd,deWit:2008ta,Bergshoeff:2008qd}.
Moreover the hierarchy is often truncated at a relatively early stage,
because the Lagrangian may be such that the gauge transformations that
connect to the higher-$p$ forms have become trivially satisfied. This
truncation process can be understood in the context of the hierarchy,
because it can be truncated (at some value of $p$) by projecting the
$p$-forms with the embedding tensor. For instance, in five space-time
dimensions, the gauged supergravity Lagrangians do not require the
presence of $p$-form fields with $p>2$, because the 2-forms appear in
the Lagrangian only in a certain contraction with the embedding tensor
that precludes the continuation to higher-$p$ gauge invariances.  On the
other hand, the $(d\!-\!1)$- and $d$-forms play a different role, as
was suggested in \cite{deWit:2008ta} where this role was explicitly
demonstrated for three-dimensional maximal supergravity.  The results
of this paper indicate that this role is in fact generic, so that,
while the hierarchy may be truncated at some specific value of $p$,
the $(d\!-\!1)$- and $d$-forms can always be included.

It is a possiblity that the truncation induced by the Lagrangian is
such that all $p$-forms with $p>1$ decouple. In that case the
hierarchy will not offer any new insights. But it does offer a
universal framework in which gaugings must take place, although the
field content of the theory will ultimately determine how much of the
hierarchal structure will be reflected in the final result. On the
other hand, the universal features of the hierarchy are presumably the
reason why the results of this approach overlap in a surprising way
with the results obtained in an entirely different context. For a
discussion of some of these results we refer to the literature (see,
e.g.  \cite{Elitzur:1997zn,Iqbal:2001ye,West:2004kb,Riccioni:2007au,
  Bergshoeff:2007qi,Riccioni:2007ni,deWit:2008ta}).

This paper is organized as follows. In section \ref{sec:hierarchy} the
hierarchy of $p$-form tensor fields is introduced in a general
context. Section \ref{sec:p-form-assignments} presents the
representations of the $p$-form fields for the maximally extended
supergravities to illustrate some of the results that can be obtained
in the context of the hierarchy. Section \ref{sec:life-at-end} deals
with the question what the generic representations are for the
higher-rank $p$-forms and discusses the possible role played by the
$(d\!-\!1)$- and $d$-form fields. In the final section some
consequences for more general deformations of ungauged theories are
pointed out.

\section{The $p$-form  hierarchy}
\setcounter{equation}{0}
\label{sec:hierarchy}

The $p$-form hierarchy has already been discussed in a number of
places, but for clarity we summarize some of its main features here.
We assume a theory with abelian gauge fields $A_\mu{}^ {M}$, that is
invariant under a group $\mathrm{G}$ of rigid transformations. The
gauge fields transform in a representation of
that group.\footnote{
  In even space-time dimensions this assignment may fail and complete
  $\mathrm{G}$ representations may require the presence of magnetic
  duals. For four space-time dimensions, this has been demonstrated in
  \cite{deWit:2005ub}.}
The generators in this representation are denoted by
$(t_\alpha)_{M}{}^{N}$, so that $\delta A_\mu{}^
{M}=-\Lambda^\alpha(t_\alpha)_{N}{}^{M} \,A_\mu{}^{N}$, and the
structure constants $f_{\alpha\beta}{}^\gamma$ of $\mathrm{G}$ are
defined according to $[t_\alpha,t_\beta]= f_{\alpha\beta}{}^\gamma
\,t_\gamma$. The next step is to select a subgroup of $\mathrm{G}$
that will be elevated to a gauge group with non-trivial gauge charges,
whose dimension is obviously restricted by the number of vector
fields.  The discussion in this section will remain rather general and
will neither depend on $\mathrm{G}$ nor on the space-time dimension.
We refer to \cite{deWit:2004nw,Samtleben:2005bp,deWit:2007mt} where a
number of results was described for maximal supergravity in various
dimensions.

The gauge group embedding is defined by specifying its generators $X_
{M}$,\footnote{
  The corresponding gauge algebra may have a central extension acting
  exclusively on the vector fields. }  
which couple to the gauge fields $A_\mu{}^ {M}$ in the usual fashion,
and which can be decomposed in terms of the independent
$\mathrm{G}$-generators $t_\alpha$, i.e.,
\begin{equation}
  \label{eq:X-theta-t}
  X_ {M} = \Theta_ {M}{}^\alpha\,t_\alpha \;.
\end{equation}
where $\Theta_ {M}{}^{\alpha}$ is the {\it embedding tensor}
transforming according to the product of the representation conjugate
to the representation in which the gauge fields transform and the
adjoint representation of $\mathrm{G}$.  This product representation
is reducible and decomposes into a number of irreducible
representations.  Only a subset of these representations is allowed.
For supergravity the precise constraints follow from the requirement
of supersymmetry, but, from all applications worked out so far, we know
that at least part (if not all) of the representation constraints is
necessary for purely bosonic reasons such as gauge invariance of the
action and consistency of the tensor gauge algebra.  This constraint
on the embedding tensor is known as the {\it representation
  constraint}. Here we treat the embedding tensor as a spurionic
object, which we allow to transform under $\mathrm{G}$, so that the
Lagrangian and transformation rules remain formally
$\mathrm{G}$-invariant. At the end we will freeze the embedding tensor
to a constant, so that the $\mathrm{G}$-invariance will be broken.  As
was shown in \cite{deWit:2008ta} this last step can also be described
in terms of a new action in which the freezing of $\Theta_
{M}{}^\alpha$ will be the result of a more dynamical process.

The embedding tensor must satisfy a second constraint, the so-called
{\em closure constraint}, which is quadratic in
$\Theta_ {M}{}^\alpha$ and more generic. This constraint
ensures that the gauge transformations form a group so that the
generators (\ref{eq:X-theta-t}) will close under commutation. Any
embedding tensor that satisfies the closure constraint, together with
the representation constraint mentioned earlier, defines a consistent
gauging. The closure constraint reads as follows,
\begin{equation}
  \label{eq:gauge-inv-embedding}
  \mathcal{Q}_{PM}{}^\alpha= 
   \Theta_{  P}{}^\beta t_{\beta  {M}}{}^{  N}
    \Theta_{  N}{}^\alpha +
    \Theta_{  P}{}^\beta f_{\beta\gamma}{}^\alpha 
    \Theta_{  M}{}^\gamma = 0 \,,
\end{equation}
and can be interpreted as the condition that the embedding tensor
should be invariant under the embedded gauge group. Hence we can write
the closure constraint as,
\begin{equation}
  \label{eq:constraint2}
  \mathcal{Q}_{MN}{}^\alpha \equiv \delta_M\Theta_N{}^\alpha=
  \Theta_M{}^\beta \,\delta_\beta\Theta_N{}^\alpha =0  \,,
\end{equation}
where $\delta_M$ and $\delta_\alpha$ denote the effect of an
infinitesimal gauge transformation or an infinitesimal
$\mathrm{G}$-transformation, respectively.  
Contracting (\ref{eq:gauge-inv-embedding}) with $t_\alpha$ leads to,
\begin{equation}
  \label{eq:XX-commutator}
  {[X_ {M},X_ {N}]} =
  -X_ {MN}{}^ {P}\,X_ {P} = 
  -X_{[ {MN}]}{}^ {P}\,X_ {P}    \,.
\end{equation}

It is noteworthy here that the generator $X_{MN}{}^{P}$ and the
structure constants of the gauge group are related, but do not
have to be identical. In particular $X_ {MN}{}^{P}$ is in general not
antisymmetric in $[{MN}]$. The embedding tensor acts as a projector,
and only in the projected subspace the matrix $X_ {MN}{}^ {P}$ is
antisymmetric in $[ {MN}]$ and the Jacobi identity will be satisfied.
Therefore (\ref{eq:XX-commutator}) implies in particular that
$X_{({MN})}{}^{P}$ must vanish when contracted with the embedding
tensor. Denoting
 \begin{equation}
  \label{eq:def-Z}
  Z^{P}{}_{{MN}} \equiv
  X_{({MN})}{}^ {P} \,,
\end{equation}
this condition reads, 
\begin{equation}
  \label{eq:closure}
  \Theta_P{}^\alpha \, Z^{P}{}_{MN} =0\,.
\end{equation}
The tensor $Z^{P}{}_{{MN}}$ is constructed by contraction of the
embedding tensor with $\mathrm{G}$-invariant tensors and therefore
transforms in the same representation as $\Theta_ {M}{}^\alpha$ ---
except when the embedding tensor transforms reducibly so that
$Z^{P}{}_{MN}$ may actually depend on a smaller representation.  The
closure constraint~(\ref{eq:constraint2}) then ensures that
$Z^{P}{}_{{MN}}$ is gauge invariant.  As is to be expected
$Z^{P}{}_{{MN}}$ characterizes the lack of closure of the generators
$X_{M}$.  This can be seen, for instance, by calculating the direct
analogue of the Jacobi identity,
\begin{equation}
\label{Jacobi-X}
   X_{[{NP}}{}^{R}\,X_{ {Q}]{R}}{}^{M} = 
   \ft23 Z^ {M}{}_{ {R}[ {N}}\,  X_{ {PQ}]}{}^ {R} \,.
\end{equation}
The fact that the right-hand side does not vanish has direct
implications for the non-abelian field strengths: the standard
expression
\begin{equation}
  \label{eq:field-strength}
  \mathcal{F}_{\mu\nu}{}^ {M} =\partial_\mu A_\nu{}^ {M}
  -\partial_\nu
  A_\mu{}^ {M} + g\, X_{[ {NP}]}{}^ {M}
  \,A_\mu{}^ {N} A_\nu{}^ {P} \,, 
\end{equation}
which appears in the commutator 
$[D_\mu,D_\nu]= - g  \mathcal{F}_{\mu\nu}{}^ {M}\,X_ {M}$
of covariant derivatives
\begin{equation}
  \label{eq:vector-gauge-tr}
 D_\mu ~\equiv~ \partial_\mu -g\, A_\mu{}^ {M}
  \,X_ {M} \,, 
\end{equation}
is not fully covariant. Rather, under standard gauge transformations
\begin{equation}
  \label{eq:A-var}
  \delta A_\mu{}^ {M} =  D_\mu\Lambda^ {M} =
  \partial_\mu \Lambda^ {M} + g A_\mu{}^ {N}
  X_ {NP}{}^ {M} \Lambda^ {P} \,,   
\end{equation}
the field strength $\mathcal{F}_{\mu\nu}{}^M$ transforms as
\begin{eqnarray}
  \label{eq:delta-cal-F}
  \delta\mathcal{F}_{\mu\nu}{}^ {M}
  &=&2\, D_{[\mu}\delta
  A_{\nu]}{}^ {M} - 
  2 g\, Z^{M}{}_{PQ} \,A_{[\mu}{}^{P}
  \,\delta A_{\nu]}{}^ {Q} \nonumber\\[1ex]
  &=&  g\, \Lambda^ {P}
  X_ {NP}{}^ {M}
  \, \mathcal{F}_{\mu\nu}{}^ {N} - 2 g\, Z^ {M}{}_ {PQ}
  \,A_{[\mu}{}^ {P}\,\delta A_{\nu]}{}^ {Q}  \,. 
\end{eqnarray} 
This expression is {\em not} covariant --- not only because of the
presence of the second term on the right-hand side, but also because
the lack of antisymmetry of $X_ {NP}{}^ {M}$ prevents us from
obtaining the expected result by inverting the order of indices ${NP}$
in the first term on the right-hand side.  As a consequence, we cannot
use $\mathcal{F}_{\mu\nu}{}^{M}$ in the Lagrangian. In particular, one
needs suitable covariant field strengths for the invariant kinetic
term of the gauge fields.

To remedy this lack of covariance, 
the strategy followed in~\cite{deWit:2004nw,deWit:2005hv}
has been to introduce additional (shift) gauge transformations 
on the vector fields, 
\begin{equation}
  \label{eq:A-var-2}
\delta A_\mu{}^ {M} =  D_\mu\Lambda^ {M} -
g\,Z^ {M}{}_{ {NP}}\,\Xi_\mu{}^{ {NP}}\,,   
\end{equation}
where the transformations proportional to $\Xi_\mu{}^ {NP}$ enable one
to gauge away those vector fields that are in the sector of the gauge
generators $X_ {MN}{}^ {P}$ in which the Jacobi identity is not
satisfied (this sector is perpendicular to the embedding tensor by
(\ref{eq:closure})).  Fully covariant field strengths can then be
defined upon introducing 2-form tensor fields $B_{\mu\nu}{}^ {NP}$
belonging to the same representation as $\Xi_\mu{}^{ {NP}}$,
\begin{equation}
  \label{eq:modified-fs}
{\cal H}_{\mu\nu}{}^{ {M}} = 
{\cal F}_{\mu\nu}{}^ {M}  + g\, Z^ {M}{}_ {NP} 
\,B_{\mu\nu}{}^ {NP}\;.
\end{equation}
These tensors transform covariantly under 
gauge transformations
\begin{eqnarray}
\delta\,\mathcal{H}_{\mu\nu}{}^{{M}}
&=&
 -g\Lambda^{P}
  X_{ {P}{N}}{}^{M}
  \mathcal{H}_{\mu\nu}{}^ {N} \,,
\end{eqnarray}
provided we impose the following transformation laws
for the 2-forms
\begin{equation}
  \label{eq:cov-delta-B}
  Z^M{}_{NP}\; \delta B_{\mu\nu}{}^{{NP}} 
  ~=~ Z^M{}_{NP}\,\Big(
   2\,D_{[\mu}\Xi_{\nu]}{}^{ {NP}} -2\, 
  \Lambda^{ {N}}\mathcal{H}_{\mu\nu}{}^{ {P}}
   + 2\, A_{[\mu}{}^{ {N}}
  \delta A_{\nu]}{}^{{P}} \Big) \,.
\end{equation}
We note that the constraint (\ref{eq:closure}) ensures that
\begin{equation}
{}
[D_\mu\,,D_\nu]~=~ - g  \mathcal{F}_{\mu\nu}{}^ {M} X_ {M}
~=~ - g  \mathcal{H}_{\mu\nu}{}^ {M} X_ {M}
\;,
\end{equation}
but in the Lagrangian the difference between $\mathcal{F}^M$
and~$\mathcal{H}^M$ is important.

Consistency of the gauge algebra thus requires the introduction of
2-form tensor fields $B_{\mu\nu}{}^ {PN}$. It is important that their
appearance in (\ref{eq:modified-fs}) strongly restricts their possible
representation content. Not only must they transform in the symmetric
product $(NP)$ of the vector field representation as is manifest from
their index structure, but also they appear under contraction with the
tensor $Z^ {M}{}_{ {NP}}$ which in general does not map onto the full
symmetric tensor product in its lower indices, but rather only on a
restricted sub-representation. It is this sub-representation of ${\rm
  G}$ to which the 2-forms are assigned, and to keep the notation
transparent, we denote the corresponding projector with special
brackets $\llceil{ {NP}}\rrfloor$, such that
\begin{equation}
Z^ {M}{}_{ {NP}} \,B_{\mu\nu}{}^{NP} ~=~
Z^ {M}{}_{ {NP}} \,B_{\mu\nu}{}^ {\llceil{ {NP}}\rrfloor}\;,\qquad
\mbox{etc.}\;.
\end{equation}
The tensor $Z^{M}{}_{{NP}}$ thus plays the role of an intertwiner
between vector fields and 2-forms, which encodes the precise field
content of the 2-form tensor fields such that the consistency of the
vector gauge algebra is ensured.

The same pattern continues upon definition of a covariant field
strength for the 2-forms and leads to a hierarchy of $p$-form tensor
fields, which is entirely determined by choice of the global symmetry
group ${\rm G}$ and its fundamental representation ${\cal R}_{\rm v}$
in which the vector fields transform.
Let us collect its main features which have emerged in the study of
particular gaugings and have been analyzed systematically
in~\cite{deWit:2005hv,deWit:2008ta}\,:
\begin{itemize}
\item
  Under the global symmetry group ${\rm G}$ of the theory, the
  $p$-forms transform in a sub-representation of the $p$-fold tensor
  product ${\cal R}_{\rm v}^{\otimes p}$, where ${\cal R}_{\rm v}$
  denotes the representation of ${\rm G}$ in which the vector fields
  transform.In many cases of interest this is the fundamental
  representation. We denote these fields by
\begin{eqnarray}
\buildrel{\scriptscriptstyle[1]}\over{A}{}^{M\vphantom{]}}\,,\quad
\buildrel{\scriptscriptstyle[2]}\over{B}{}^{{\llceil MN\rrfloor}}\,,\quad
\buildrel{\scriptscriptstyle[3]}\over{C}{}^{{\llceil M\llceil
    NP\rrfloor\rrfloor}}\,,\quad 
\buildrel{\scriptscriptstyle[4]}\over{C}{}^{{\llceil M\llceil N\llceil
    PQ\rrfloor\rrfloor\rrfloor}}\,,\quad 
\buildrel{\scriptscriptstyle[5]}\over{C}{}^{{\llceil M\llceil N\llceil
    P\llceil{QR}\rrfloor\cdot\cdot\rrfloor}}\,,\;\; {\rm etc.}\,,
\end{eqnarray}
where we have suppressed space-time indices, and the
special brackets $\llceil\cdots\rrfloor$ are introduced to denote the
relevant sub-representations of ${\cal R}_{\rm v}^{\otimes p}$.

\item
The precise representation content of the $(p+1)$-forms 
$C_{[p+1]}{}^{\llceil{N}_0\llceil{N}_1\llceil
  \cdots {N}_p\rrfloor\cdot\cdot\rrfloor}$
are reflected in the {\em intertwining tensors} $Y$, defined
recursively in terms of the lower-rank intertwiners and  
the gauge group generators $X_{N_0}$ evaluated in the 
representation of the $p$-forms. For $p\ge3$, this recursive relation
is given by 
\begin{eqnarray}
  \label{eq:higherY}
  Y^{{M}_1\llceil{M}_2\llceil
  \cdots {M}_p\rrfloor\cdot\cdot\rrfloor}  
     {}_{{N}_0\llceil{N}_1\llceil
  \cdots {N}_p\rrfloor\cdot\cdot\rrfloor}   
   &\equiv&{} 
   - \delta^{\llceil {M}_1}_{{N}_0}\;
   Y_{\vphantom{\llceil}}^{{M}_2\llceil
  \cdots {M}_p\rrfloor\cdot\cdot\rrfloor}
  {}^{\vphantom{\llceil}}_{{N}_1\llceil 
  {N}_2\llceil\cdots\, {N}_p\rrfloor\cdot\cdot\rrfloor} 
   \nonumber\\[1.5ex]
    &&{}
    - (X_{ {N}_0})_{\llceil {N}_1\llceil {N}_2\llceil\cdots\,
     {N}_p\rrfloor\cdot\cdot\rrfloor}
  {}^{\llceil {M}_1\llceil {M}_2\llceil\cdots\,
     {M}_p\rrfloor\cdot\cdot\rrfloor}\;\;. 
\end{eqnarray}
Inspection of (\ref{eq:higherY}) for a concrete choice of ${\rm G}$
and ${\cal R}_{\rm v}$ shows that the intertwining tensor, considered
as a map
\begin{eqnarray}
Y^{[p]}:\,
{\cal R}_{\rm v}^{\otimes (p+1)}\;\longrightarrow\;{\cal R}_{\rm v}^{\otimes p}
\;,
  \label{eq:map}
\end{eqnarray}
has a non-trivial kernel whose complement 
defines the representation content of the $(p+1)$-forms
that is required for consistency of the deformed $p$-form gauge algebra.

It is important to stress that all intertwining tensors depend
linearly on the embedding tensor~$\Theta$. Since they are constructed
from the embedding tensor contracted with $\mathrm{G}$-invariant
tensors, they all transform covariantly and belong to the same
representation as the embedding tensor, in spite of their different
index structure.  Obviously the intertwining tensors depend on the
particular gauging considered.  However, sweeping out the full space
of possible embedding tensors yields a $\Theta$-independent (and ${\rm
  G}$-covariant) result for the representation of $(p+1)$-forms. This
is understood by regarding the embedding tensor as a so-called {\it
  spurionic} quantity, which transforms under the action of
$\mathrm{G}$, although at the end it will be fixed to a constant
value. This approach shows how the mere consistency of the deformation
of the $p$-form gauge algebra upon generic gaugings imposes rather
strong restrictions on the field content of the ungauged theory.  In
the ungauged theory there is a priori no direct evidence for these
restrictions and usually additional structures, such as supersymmetry
or the underlying higher-rank Kac-Moody symmetries, motivate the
presence and precise field content of the $p$-forms. It is rather
surprising and intriguing that the constraints implied by these
additional structures on the field content do precisely coincide with
the constraints derived from consistency of the $p$-form hierarchy.
\item
The lowest-rank intertwining tensors are given by 
\begin{eqnarray}
  \label{eq:lowerY}
Y^{[0]}:\,
{\cal R}_{\rm v}\;\longrightarrow\;{\cal R}_{\rm adj}
\;,\qquad
Y^{[1]}:\,
{\cal R}_{\rm v}^{\otimes 2}\;\longrightarrow\;{\cal R}_{\rm v}
\;,
\end{eqnarray}
corresponding to $p=0,1$, with
$(Y^{[0]})^\alpha{}_{M}=\Theta_M{}^\alpha$ and
$(Y^{[1]})^M{}_{PQ}=Z^M{}_{PQ}$. For $p=2$, the intertwining tensor
can be written as follows,
\begin{eqnarray}
  \label{eq:def-Y-2}
  Y^{MN}{}_{{P}\llceil {RS}\rrfloor} =
  2 \,\delta_ {P}{}^{\llceil {M}}
  \,Z^{ {N}\rrfloor}{}_ {RS} 
  -  X_{P\llceil {RS}\rrfloor}{}^{\llceil MN\rrfloor} \;.
\end{eqnarray}

\item
  Inspection of the symmetry properties of the intertwining tensors
  (\ref{eq:lowerY}) and (\ref{eq:higherY}) shows explicitly that in
  general the lowest-rank $p$-forms in the hierarchy do not live in
  the full tensor product ${\cal R}_{\rm v}^{\otimes p}$, but only in a
  subsector thereof constrained by certain symmetry properties:
\begin{eqnarray}
  \buildrel{\scriptscriptstyle[1]}\over{A}\;\, \in \tinyyoung{\cr}
\;,\qquad
 \buildrel{\scriptscriptstyle[2]}\over{B}\;\,  \in \tinyyoung{&\cr}
\;,\qquad
 \buildrel{\scriptscriptstyle[3]}\over{C}\;\,  \in \tinyyoung{&\cr\cr}
\;,\qquad
 \buildrel{\scriptscriptstyle[4]}\over{C}\;\,  \in
 \tinyyoung{&&\cr\cr}\oplus\tinyyoung{&\cr\cr\cr}  \;,\;\;
{\rm etc.}\,,
\end{eqnarray}
in standard Young tableau notation.\footnote{We should stress that the
  Young box `$\tinyyoung{\cr}$\,' here corresponds to the
  representation ${\cal R}_{\rm v}$ in which the vector fields
  transform under ${\rm G}$ and {\em not} to their space-time
  structure.  With respect to the latter, all tensors of course
  transform as $p$-forms, i.e.\ in the totally antisymmetric part of
  the $p$-fold tensor product.} In general, the group ${\rm G}$ will
be different from an~${\rm SL}(N)$, so that the Young tableaux
themselves are reducible. As it turns out, the tensor hierarchy then
imposes further restrictions on the representation content.

\item
Mutual orthogonality: the intertwining tensors satisfy the relations
\begin{eqnarray}
  \label{eq:orthoY}
  Y^{{K}_2\llceil{K}_3\llceil\cdots {K}_p\rrfloor
  \cdot\cdot\rrfloor}
  {}_{{M}_1\llceil {M}_2\llceil\cdots
   {M}_p\rrfloor\cdot\cdot\rrfloor} \; 
   Y^{{M}_1\llceil {M}_2\llceil\cdots
   {M}_p\rrfloor\cdot\cdot\rrfloor}
  {}_{{N}_0\llceil{N}_1\llceil\cdots
   {N}_p\rrfloor\cdot\cdot\rrfloor}  
   &\approx&{} 0 \;,
\end{eqnarray}
where `weakly zero' ($\approx 0$) indicates that the expression
vanishes as a consequence of the quadratic constraint
(\ref{eq:gauge-inv-embedding}) on the embedding tensor. More
schematically, these orthogonality relations take the form
\begin{eqnarray}
  \label{eq:orthoY1}
  Y^{[p]}  \cdot Y^{[p+1]} &\approx& 0
  \;,
\end{eqnarray}
(with equation (\ref{eq:closure}) as their lowest member)
and thus in view of (\ref{eq:map}) define the sequence
\begin{eqnarray}
\cdots\;
\stackrel{Y^{[p+1]}}{\longrightarrow}\;
{\cal R}_{\rm v}^{\otimes (p+1)}\;
\stackrel{Y^{[p]}}{\longrightarrow}\;
{\cal R}_{\rm v}^{\otimes p}\;
\stackrel{Y^{[p-1]}}{\longrightarrow}\;
\;\;\;\cdots\;\;\;
\stackrel{Y^{[1]}}{\longrightarrow}\;
{\cal R}_{\rm v}\;\stackrel{Y^{[0]}}{\longrightarrow}\;{\cal R}_{\rm adj}
\;.
\label{sequence}
\end{eqnarray}
Again, we emphasize that every embedding tensor, i.e.\ every solution
to the quadratic constraint, gives rise to such a sequence and defines
its proper field content, while by sweeping out the entire space of
possible embedding tensors one obtains the full $p$-form field content
induced by the group ${\rm G}$.
   
\item
  Consequently, given the $Y$-tensors, and specifying the group G, the
  above results enable a complete determination of the full hierarchy
  of the higher-rank $p$-forms required for the consistency of the
  gauging. In particular, we can exhibit some of the terms in the
  variations of the $p$-form fields
\begin{eqnarray}
  \label{eq:trans-C-p}
  \delta   \buildrel{\scriptscriptstyle[p]}\over{C}{}^{ {M}_1\llceil
   {M}_2\llceil\cdots {M}_p\rrfloor\cdot\cdot\rrfloor}
  &=&{}
  p\,{\mathrm D}\!   \buildrel{\scriptscriptstyle[p-1]}\over{\Phi}{}^{ {M}_1\llceil
   {M}_2\llceil\cdots {M}_p\rrfloor\cdot\cdot\rrfloor} 
   \nonumber \\[1.5ex]
   &&{}
   + \Lambda^{\llceil {M}_1}
       \buildrel{\scriptscriptstyle[p]}\over{{\mathcal{H}}}{}^{\llceil {M}_2
  \cdots\rrfloor\cdot\cdot\rrfloor} 
   +p\,\delta   \buildrel{\scriptscriptstyle[1]}\over{A}{}^{\llceil {M}_1}\wedge
    \,  \buildrel{\scriptscriptstyle[p-1]}\over{C}{}^{\llceil {M}_2
  \cdots\rrfloor\cdot\cdot\rrfloor}  
    \nonumber\\[1.5ex]
   &&{} 
   -g\,Y^{{M}_1\llceil{M}_2\llceil\cdots
   {M}_p\rrfloor\cdot\cdot\rrfloor}  
   {}_{{N}_0\llceil {N}_1\llceil
  \dots {N}_p\rrfloor.\rrfloor} \; 
   \buildrel{\scriptscriptstyle[p]}\over{\Phi}
   {}^{{N}_0\llceil{N}_1\llceil 
  \dots {N}_p\rrfloor\cdot\cdot\rrfloor}\;,  
   \nonumber\\[1.5ex]
   &&{}+  \cdots\;. 
\end{eqnarray} 
In particular, this demonstrates how the intertwining tensors $Y$ show
up explicitly in the tensor gauge transformations to induce a
St\"uckelberg-type coupling between $p$ and $(p+1)$-forms.  The dots
in (\ref{eq:trans-C-p}) represent further terms carrying the
lower-rank $p$-forms such as terms linear in the covariant field
strengths $\mathcal{H}$ (to be introduced below) and further
Chern-Simons-like variations such as $\delta C \wedge C$.

\item
  For all higher-rank $p$-forms covariant field strengths can be
  defined that transform homogeneously under vector gauge
  transformations and are invariant under all higher-rank tensor gauge
  transformations.\\
  E.g.\ for the 2-forms the modified field strength takes the form,
\begin{eqnarray}
  \label{eq:H-3-cov}
    \buildrel{\scriptscriptstyle[3]}\over{\mathcal{H}}{}^{{MN}}&\equiv&
   3 \,{\mathrm D}\! \buildrel{\scriptscriptstyle[2]}\over{B}{}^{MN}  +
  3\,
\buildrel{\scriptscriptstyle[1]}\over{A}{}^{\llceil{M}}\wedge
  \left(\mathrm{d}\buildrel{\scriptscriptstyle[1]}\over{A}{}^{{N}\rrfloor}
  + \ft23 g   X_{[{PQ}]}{}^{{N}\rrfloor}
    \buildrel{\scriptscriptstyle[1]}\over{A}{}^{P}\!\wedge\!
  \buildrel{\scriptscriptstyle[1]}\over{A}{}^{Q}\right)  
  \nonumber\\
  &&{}   + g\,
  Y^{MN}{}_{{P}\llceil{RS}\rrfloor}
  \;  \buildrel{\scriptscriptstyle[3]}\over{C}{}^{{P}\llceil{RS}\rrfloor} \;.
\end{eqnarray}
This pattern continues. 

\item
  The hierarchy can be truncated at any value of $p$ by projecting the
  corresponding forms with the next intertwining tensor. Because of
  the orthogonality property (\ref{eq:orthoY}), the St\"uckelberg-type
  shifts are then no longer effective and the hierarchy will not be
  continued to higher $p$-forms. Of course, this projection is a
  somewhat arbitrary and technical way to truncate, but in practice
  this situation may occur when considering specific Lagrangians in
  which intertwining tensors may appear that effect precisely this
  projection. For instance, in five-dimensional maximal supergravity,
  the 3-form fields do not appear in the Lagrangian for precisely this
  reason.

\end{itemize}

Although the number of space-time dimensions does not enter into this 
analysis (as stated earlier, the iteration procedure can in principle be 
continued indefinitely), there exists, for the maximal supergravities, a
consistent correlation between the rank of the tensor fields and the
occurrence of conjugate $\mathrm{G}$-representations that is precisely
in accord with tensor-tensor and vector-tensor (Hodge)
duality\footnote{
  As well as with the count of physical degrees of freedom.}
corresponding to the space-time dimension where the maximal
supergravity with that particular duality group $\mathrm{G}$ lives. In
the next section we discuss some of the results of this analysis.

\section{Representation assignments of the $p$-forms}
\setcounter{equation}{0}
\label{sec:p-form-assignments}
The hierarchy of vector and tensor gauge fields that we presented in
the previous section can be considered in the context of the maximal
gauged supergravities. In that case the gauge group is embedded in the
duality group $\mathrm{G}$, which is known for each space-time
dimension in which the supergravity is defined. Once the group
$\mathrm{G}$ is specified, the hierarchy allows in principle a unique
determination of the representations of the higher $p$-forms. Table
\ref{tab:vector-tensor-repr} shows an overview of some of the results.
We recall that the analysis described in section~\ref{sec:hierarchy}
did not depend on the number of space-time dimensions. For instance,
it is possible to derive the representation assignments for
$(d\!+\!1)$-rank tensors, although these do not live in a
$d$-dimensional space-time (nevertheless, a glimpse of their existence
occurs in $d$ dimensions via the shift transformations
(\ref{eq:trans-C-p}) of the $d$-forms in the general gauged theory).

On the other hand, whenever there exists a (Hodge) duality relation
between fields of different rank at the appropriate value for $d$,
then one finds that their $\mathrm{G}$ representations turn out to be
related by conjugation. This property is clearly exhibited at the
level of the lower-rank fields in the table.  More precisely, upon
working out the precise representation content as described in the
previous section, the sequence (\ref{sequence}) takes the particular
form
\begin{eqnarray}
\cdots\;
\stackrel{Y^{[d-2]}}{\longrightarrow}\;
{\cal R}_{\rm adj}\;
\stackrel{Y^{[d-3]}}{\longrightarrow}\;
{\cal R}_{\rm v*}\;
\stackrel{Y^{[d-4]}}{\longrightarrow}\;
\;\;\;\cdots\;\;\;
\stackrel{Y^{[1]}}{\longrightarrow}\;
{\cal R}_{\rm v}\;\stackrel{Y^{[0]}}{\longrightarrow}\;{\cal R}_{\rm adj}
\;,
\label{sequence-symmetric}
\end{eqnarray}
symmetric around the forms of rank $p=\frac12[d\!-\!1]$, i.e.\ ${\cal
  R}_{\rm v*}$ denotes the representation dual to~${\cal R}_{\rm v}$,
etc..  In particular, the intertwiners in (\ref{sequence-symmetric})
are pairwise related by transposition
\begin{eqnarray}
Y^{[0]}=(Y^{[d-3]}\,)^{\rm T}\;,\qquad 
Y^{[1]}=(Y^{[d-4]}\,)^{\rm T}\;,\qquad
\mbox{etc.}
\;.
\end{eqnarray}
It is intriguing that the purely group theoretical hierarchy reproduces
the correct assignments consistent with Hodge duality.
In particular, the assignment of the $(d\!-\!2)$-forms is in line with 
tensor-scalar duality, as these
forms are dual to the Noether currents associated with the
$\mathrm{G}$ symmetry.
In this sense, the duality group $\mathrm{G}$ implicitly carries 
information about the space-time dimension.

What is more, the hierarchy naturally extends beyond the
$(d\!-\!2)$-forms and thus to those non-propagating forms whose field
content is not restricted by Hodge duality.  It is another striking
feature of the hierarchy that the diagonals pertaining to the
$(d\!-\!1)$- and $d$-rank tensor fields refer to the representations
conjugate to those assigned to the embedding tensor and its quadratic
constraint, respectively.  In the next section, we will show that this
pattern is in fact generic and related to the special role these forms
may play in the Lagrangian~\cite{deWit:2008ta}.

\begin{table}[t]
\centering
\begin{tabular}{l l cccccc  }\hline
~&~&~&~&~&~&~& \\[-4mm]
~ &$~$& 1&2&3&4&5&6  \\   \hline
~&~&~&~\\[-4mm]
7   & ${\rm SL}(5)$ & $\overline{\bf 10}$  & ${\bf 5}$ & $\overline{\bf 5}$ &
${\bf 10}$ &  ${\bf 24}$ & $\overline{\bf 15}+{\bf 40}$  \\[1mm]
6  & ${\rm SO}(5,5)$ & ${\bf 16}_c$ & ${\bf 10}$ & ${\bf 16}_s$ & 
  ${\bf 45}$ & ${\bf 144}_s$ &  
$\!\!{\bf 10}\!+\! {\bf 126}_s\!+\! {\bf 320}\!\!$\\[.8mm]
5   & ${\rm E}_{6(6)}$ & $\overline{\bf 27}$ & ${\bf 27}$ & ${\bf 78}$
& ${\bf 351}$ & $\!\!{\bf 27}\!+\! {\bf 1728}\!\!$ &  \\[.5mm]
4   & ${\rm E}_{7(7)}$ & ${\bf 56}$ & ${\bf 133}$ & ${\bf 912}$ &
$\!\!\!{\bf 133}\!+\! {\bf 8645}\!\!\!$ &    \\[.5mm]
3   & ${\rm E}_{8(8)}$ & ${\bf 248}$ & ${\bf 1}\!+\! {\bf 3875}$ & ${\bf
  3875}\!+\! {\bf147250}$ & & 
\\ \hline
\end{tabular}
\caption{\small
Duality representations of the vector and tensor gauge fields
for gauged  maximal supergravities in space-time dimensions $3\leq
d\leq 7$. The first two columns list the space-time dimension and the
corresponding duality group. }\label{tab:vector-tensor-repr} 
\end{table}

It is an obvious question whether these systematic features have a
natural explanation in terms of M-theory and we refer to
\cite{deWit:2008ta} for a discussion. Here it suffices to mention that
the representation content agrees with results based on matrix models
in M-theory \cite{Elitzur:1997zn}, (see also, \cite{Obers:1998fb} and
references quoted therein) where matrix theory
\cite{deWit:1988ig,Banks:1996vh} is considered in a toroidal
compactification. The representations in the table were also found in
\cite{Iqbal:2001ye}, where a `mysterious duality' was exhibited
between toroidal compactifications of M-theory and del Pezzo surfaces.
Here the M-theory dualities are related to global diffeomorphisms that
preserve the canonical class of the del Pezzo surface. Again the
representations thus found are in good agreement with the
representations in table~\ref{tab:vector-tensor-repr}.  Furthermore
there are hints that the above considerations concerning new
M-theoretic degrees of freedom can be extended to infinite-dimensional
duality groups. Already some time ago \cite{West:2004kb} it was shown
from an analysis of the indefinite Kac--Moody algebra
$\mathrm{E}_{11}$ that the decomposition of its so-called L1
representation at low levels under its finite-dimensional subalgebra
$\mathrm{SL}(3) \times \mathrm{E}_{8}$ yields the same $\mathbf{3875}$
representation that appears for the 2-forms as shown in
table~\ref{tab:vector-tensor-repr}. This analysis has meanwhile been
extended \cite{Riccioni:2007au,Bergshoeff:2007qi,Riccioni:2007ni} to
other space-time dimensions and higher-rank forms, and again there is
a clear overlap with the representations in
table~\ref{tab:vector-tensor-repr}. Non-maximal supergravities have
also been discussed from this perspective in
\cite{Bergshoeff:2007vb,Riccioni:2008jz}. 

\section{Life at the end of the hierarchy}
\label{sec:life-at-end}
\setcounter{equation}{0}
Historically the $p$-form hierarchy was discovered by starting from
the 1-forms belonging to the representation ${\cal R}_{\rm v}$, in the
context of specific (supergravity) theories. The crucial ingredients
are the group $\mathrm{G}$ and the representation of the embedding
tensor. No information about the space-time dimension is required. On
the other hand, one of the intitial observations was that general
gaugings require a certain decomposition between certain $p$-forms and
their duals, which belong to the conjugate representation. The actual
distribution of physical degrees of freedom over these sets of fields
related by duality is eventually determined by the value taken by the
embedding tensor.

In this section, we will study the generic representation content of
the $p$-forms predicted by the hierarchy for large rank $p$ close to
$d$.  In view of the fact that the theory is invariant under the group
$\mathrm{G}$ prior to switching on the gauge couplings, there exists a
set of conserved 1-forms given by the Noether currents, transforming
in the adjoint representation, which is dual to the $(d\!-\!2)$-forms.
Furthermore we expect $(d\!-\!3)$-forms that are dual to the vector
fields and thus are expected to transform in the ${\rm G}$
representation ${\cal R}_{\rm v*}$ dual to the vector field
representation, in accordance with~(\ref{sequence-symmetric}).  When
considering these high-rank $p$-forms it is convenient to switch from
the general notation that was used in section~\ref{sec:hierarchy} to a
notation adapted to this particular field content and to identify the
$(d\!-\!3)$- and $(d\!-\!2)$-forms as,
\begin{eqnarray}
  \label{eq:d-3-d-2} 
  \buildrel{\scriptscriptstyle[d-3]}\over{C}{}^{ {M}_1\llceil
   {M}_2\llceil\cdots {M}_{d-3}\rrfloor\cdot\cdot\rrfloor}
   &\sim & \buildrel{\scriptscriptstyle[d-3]}\over{C}_ M
   \;, \nonumber\\[.2ex] 
   \buildrel{\scriptscriptstyle[d-2]}\over{C}{}^{ {M}_1\llceil
   {M}_2\llceil\cdots {M}_{d-2}\rrfloor\cdot\cdot\rrfloor}
   &\sim & \buildrel{\scriptscriptstyle[d-2]}\over{C}_\alpha \;,
\end{eqnarray}
upon explicit introduction of corresponding projectors, denoted by
${\mathbb{P}}^{ {M}_1\llceil {M}_2\llceil\cdots
  {M}_{d-3}\rrfloor\cdot\cdot\rrfloor}{}_M$ and ${\mathbb{P}}^{
  {M}_1\llceil
  {M}_2\llceil\cdots{M}_{d-3}\rrfloor\cdot\cdot\rrfloor}{}_\alpha$. We
may then explicitly study the end of the $p$-form hierarchy by
imposing the general structure outlined in 
section~\ref{sec:hierarchy}. The result takes the following form, 
\begin{eqnarray}
  \label{eq:general-hierarchy-end}
  \delta \buildrel{\scriptscriptstyle[d-3]}\over{C}{}_M &=&
  (d-3) \,{\mathrm D}\!
  \buildrel{\scriptscriptstyle[d-4]}\over{\Phi}{}_M + \cdots  
  - Y_M{}^\alpha\,\buildrel{\scriptscriptstyle[d-3]}\over{\Phi}{}_\alpha
  \;,\nonumber\\[1.5ex]
  \delta \buildrel{\scriptscriptstyle[d-2]}\over{C}{}_\alpha &=& 
  (d-2)\,{\mathrm D}\!
  \buildrel{\scriptscriptstyle[d-3]}\over{\Phi}{}_\alpha + \cdots  
  - Y_{\alpha,M}{}^\beta
  \buildrel{\scriptscriptstyle[d-2]}\over{\Phi}{}^M{}_\beta
  \;,\nonumber\\[1.5ex]
  \delta \buildrel{\scriptscriptstyle[d-1]}\over{C}{}^M{}_\alpha &=& 
  (d-1)\,{\mathrm D}\!
  \buildrel{\scriptscriptstyle[d-2]}\over{\Phi}{}^M{}_\alpha+ \cdots  
  - Y^M{}_{\alpha, PQ}{}^\beta 
  \buildrel{\scriptscriptstyle[d-1]}\over{\Phi}{}^{PQ}{}_\beta
  \;,\nonumber\\[1.5ex]
  \delta \buildrel{\scriptscriptstyle[d]}\over{C}{}^{MN}{}_\alpha &=&
  d\,{\mathrm D}\!
  \buildrel{\scriptscriptstyle[d-1]}\over{\Phi}{}^{MN}{}_\alpha 
  + \cdots - Y^{MN}{}_{\alpha, PQR}{}^\beta \,
  \buildrel{\scriptscriptstyle[d]}\over{\Phi}{}^{PQR}{}_\beta
  \;,\nonumber\\[1.5ex] 
  \delta \buildrel{\scriptscriptstyle[d+1]}\over{C}{}^{PQR}{}_\alpha &=&
  (d+1)\,{\mathrm D}\!
  \buildrel{\scriptscriptstyle[d]}\over{\Phi}{}^{PQR}{}_\alpha 
  + \cdots  \,,
\end{eqnarray}
where we indicated the most conspicuous parts of the $p$-form
transformations. We included the transformations associated to the
$(d\!+\!1)$-form for reasons that will be explained shortly.

From the index structure it is obvious that $Y_M{}^\alpha$ must
coincide with the embedding tensor. The subsequent
intertwining tensors can then be found by applying~(\ref{eq:higherY})
which yields\footnote{
  It is important to realize that (\ref{eq:higherY}) is only valid for
  $p\geq3$, which implies that these results cannot be directly applied
  to low space-time dimensions. However, in that case the intertwining
  tensors are already known and given by (\ref{eq:lowerY}) and
  (\ref{eq:def-Y-2}).} 
\begin{eqnarray}
  \label{eq:interpolater-consistency}
  Y_{\alpha,M}{}^\beta &=& t_{\alpha M}{}^N\,Y_N{}^\beta -
  X_M{}^\beta{}_\alpha \;,
  \nonumber\\[1.2ex]
  Y^M{}_{\alpha, PQ}{}^\beta  &=&{}- \delta_P{}^M\,Y_{\alpha,Q}{}^\beta
  -(X_P)_Q{}^{\beta,M}{}_\alpha \;,
  \nonumber\\[1.2ex]
  Y^{MN}{}_{\alpha, PQR}{}^\beta  &=&{}
  -\delta^M_P\,Y^N{}_{\alpha, QR}{}^\beta -
  (X_P)_{QR}{}^{\beta,MN}{}_\alpha    \;.
\end{eqnarray}
The presence of the generator $t_{\alpha M}{}^N$ in the first equation
is related to the conversion of the special bracket notation employed
in the previous sections.

It is, however, more instructive to cast these expressions into a
different form, given by
\begin{eqnarray}
  \label{eq:interpolaters}
  Y_M{}^\alpha &=& 
  \Theta_M{}^\alpha \;,
  \nonumber\\[1.5ex]
  Y_{\alpha,M}{}^\beta &=&
  \delta_\alpha \Theta_M{}^\beta\;,
  \nonumber\\[1.5ex]
  Y^M{}_{\alpha, PQ}{}^\beta  &=&{}- 
  \frac{\partial{\cal Q}_{PQ}{}^\beta}{\partial\,\Theta_M{}^\alpha} \;,
  \nonumber\\[1.5ex]
  Y^{MN}{}_{\alpha, PQR}{}^\beta  &=&
  -\delta^M_P\,Y^N{}_{\alpha, QR}{}^\beta
  - X_{PQ}{}^M\,\delta_R^N\delta_\alpha^\beta
  - X_{PR}{}^N\,\delta_Q^M \delta_\alpha^\beta
  + X_{P\alpha}{}^\beta\,\delta_R^N\delta_Q^M 
  \;,
\end{eqnarray}
where sign factors have been adopted such that the above tensors are
precisely consistent with (\ref{eq:interpolater-consistency}). In this
form, it is straightforward to verify that the intertwining tensors
satisfy the mutual orthogonality property (\ref{eq:orthoY}). For the
first few tensors this is easy to prove,
\begin{eqnarray}
  \label{eq:YY-aprox-Q}
  Y_M{}^\alpha \,Y_{\alpha,N}{}^\beta &=&
  \delta_M \Theta_N{}^\beta ~=~ {\cal Q}_{MN}{}^\beta~\approx~0  \;,
  \nonumber\\[1ex]
  Y_{\alpha,N}{}^\beta\,Y^N{}_{\beta, PQ}{}^\gamma &=&
  \delta_\alpha  \,{\cal Q}_{PQ}{}^\gamma~\approx~0 \;,
\end{eqnarray}
where we recall the constraint written as in (\ref{eq:constraint2}).
In the second equation we used the fact that the intertwining tensors
are all $\mathrm{G}$-covariant, so that the effect of transforming the
embedding tensor is equivalent to transforming the tensor according to
its index structure.

The last orthogonality relation is proved differently. First we
note the identity,
\begin{equation}
  \label{eq:Y-Q-vanish}
  Y^{MN}{}_{\alpha, PQR}{}^\beta \; {\cal Q}_{MN}{}^\alpha
  = 0 \;,  
\end{equation}
which holds identically without making reference to the quadratic
constraint~(\ref{eq:gauge-inv-embedding}). This is thus a non-trivial
identity that is cubic in the embedding tensor. It follows by
comparing 
\begin{equation}
  \label{eq:delta-Q1}
  \delta_P\, {\cal Q}_{QR}{}^\beta  =
  \delta_P\Theta_N{}^\alpha\,
  \frac{\partial{\cal Q}_{QR}{}^\beta}{\partial\,{\Theta_N{}^\alpha}}
  = \Big(- \delta^M_P\,Y^N{}_{\alpha, QR}{}^\beta\Big)\, {\cal
  Q}_{MN}{}^\alpha  \;,
\end{equation}
to 
\begin{equation}
  \label{eq:delta-Q2}
  \delta_P\, {\cal Q}_{QR}{}^\beta = 
  \Big(X_{PQ}{}^M\,\delta_R^N\delta_\alpha^\beta 
  + X_{PR}{}^N\,\delta_Q^M \delta_\alpha^\beta
  - X_{P\alpha}{}^\beta\,\delta_R^N\delta_Q^M 
  \Big)\, {\cal Q}_{MN}{}^\alpha   \;.
\end{equation}
This last equation follows from the fact that the tensor
$\mathcal{Q}_{QR}{}^\beta$ transforms covariantly. Taking the
difference of the two equations (\ref{eq:delta-Q1}) and
(\ref{eq:delta-Q2}) leads directly to (\ref{eq:Y-Q-vanish}). 

The importance of this result will be discussed below, but we first
note that the missing orthogonality relation between the intertwiners
follows from taking the derivative of (\ref{eq:Y-Q-vanish}) with
respect to the embedding tensor,
\begin{eqnarray}
  \label{eq:YY-final}
  Y^M{}_{\alpha, KL}{}^\beta\,
  Y^{KL}{}_{\beta, PQR}{}^\gamma   &=&{}
  - Y^{KL}{}_{\beta, PQR}{}^\gamma \;
  \frac{\partial{\cal Q}_{KL}{}^\beta}{\partial\,\Theta_M{}^\alpha}
  \nonumber\\ 
  &=&{}
  \frac{\partial Y^{KL}{}_{\beta, PQR}{}^\gamma} 
  {\partial\,\Theta_M{}^\alpha} \;{\cal Q}_{KL}{}^\beta
  ~\approx~0  \;.   
\end{eqnarray}

From~(\ref{eq:interpolaters}) we can now directly read off the
representation content of the $(d\!-\!1)$- and the $d$-forms that
follows from the hierarchy: the form of $Y_{\alpha,M}{}^\beta$ and
$Y^M{}_{\alpha, PQ}{}^\beta$ shows that these forms transform in the
representations dual to the embedding tensor $\Theta_M{}^\beta$ and
the quadratic constraint ${\cal Q}_{PQ}{}^\beta$, respectively.  As
such, they can naturally be coupled, acting as Lagrange multipliers
enforcing the property that the embedding tensor is space-time
independent and gauge invariant \cite{deWit:2008ta}.  This idea has
been worked out explicitly in the context of maximal supergravity in
three space-time dimensions, and we will demonstrate here that it can
also be realized in a more general context. Hence we view the
embeddding tensor as a space-time dependent scalar field, transforming
in the $\mathrm{G}$-representation constrained by possible
representation constraints. To the original Lagrangian $\mathcal{L}_0$
which may depend on $p$-forms with $p\leq d-2$, we then add the
following interactions,
\begin{equation}
  \label{eq:addtion}
  \mathcal{L} = \mathcal{L}_0+\mathcal{L}_\mathrm{C}\,,
\end{equation}
with 
\begin{eqnarray}
  \label{eq:LC}
  \mathcal{L}_\mathrm{C} &\propto&{} 
  \varepsilon^{\mu_1\cdots\mu_d}\left\{ d\,g \,  
  C_{\mu_2\cdots\mu_d}{}^{M}{}_\alpha \,D_{\mu_1}\Theta_M{}^\alpha  
  + g^{2}\,
  C_{\mu_1\cdots\mu_d}{}^{MN}{}_\alpha\;
  \mathcal{Q}_{MN}{}^\alpha\right\}  \;,
\end{eqnarray}
where $\Theta_M{}^\alpha(x)$ is now a field. First we note that this
Lagrangian is invariant under the shift transformation of the $d$-rank
tensor field, by virtue of the identity (\ref{eq:Y-Q-vanish}). Varying
this Lagrangian with respect to $\Theta_M{}^\alpha$ leads to the
following variation,
\begin{eqnarray}
  \label{eq:delta-LC}
  \delta\mathcal{L}_\mathrm{C} &\propto& -
  g\, \varepsilon^{\mu_1\cdots\mu_d} \;\delta\Theta_M{}^\alpha
  \nonumber \\[.5ex] 
  &&{} \quad\times
  \Big[d\,D_{\mu_1}C_{\mu_2\cdots\mu_d}{}^M{}_\alpha  +g\, 
  Y^M{}_{\alpha,PQ}{}^\beta\, 
  C_{\mu_1\cdots\mu_d}{}^{PQ}{}_\beta  
  +d\, g\, A_{\mu_1}\, Y_{\alpha,N}{}^\beta\;
  C_{\mu_2\cdots\mu_d}{}^N{}_\beta \Big] 
  \,.\nonumber\\[.5ex]
 \end{eqnarray}
This result can be written as follows,
\begin{equation}
  \label{eq:vra-into-H}
  \delta\mathcal{L}_\mathrm{C}\propto -
  g\,\varepsilon^{\mu_1\cdots\mu_d}\,\Big[
  \mathcal{H}_{\mu_1\cdots\mu_d}{}^M{}_\alpha + d\, g A_{[\mu_1}{}^M
  \;\mathcal{H}_{\mu_2\cdots\mu_d]\;\alpha} +\cdots\Big]
  \,\delta\Theta_M{}^\alpha \,, 
\end{equation}
by including unspecified terms involving form fields of rank $p\leq
d-2$. These terms are assumed to originate from the $\Theta$-variation
of the Lagrangian $\mathcal{L}_0$, but they cannot be evaluated in
full generality as this depends on the details of the latter Lagrangian.

Qualitatively the above result is quite similar to that obtained in
three space-time dimensions, but there are slight differences in the
numerical factors, due to the fact that the three-dimensional result
involves the intertwining tensors for low $p$-values, whereas the
result here is based on generic $p\geq3$ intertwining tensors. For
Lagrangians that contain at most two derivatives, the Lagrangian will
depend at most quadratically on $\Theta_M{}^\alpha(x)$. Hence this
field may be integrated out, precisely as discussed in three
space-time dimensions \cite{deWit:2008ta}, so that all possible
gaugings are comprised in one single Lagrangian.
\section{Concluding remarks}
\label{sec:conclusions}
\setcounter{equation}{0}

The gaugings accompanied by a $p$-form hierarchy can be considered as
a class of deformations of the original theory (which was invariant
under the group $\mathrm{G}$), induced by switching on certain
charges. These charges necessarily generate a subgroup of
$\mathrm{G}$, extended by a variety of $p$-form gauge transformations.
In principle, other deformations can be envisaged and one may
wonder whether they can be switched on at the same time and/or whether
they are completely independent.

An example constitutes the massive deformation known from IIA
supergravity in ten dimensions~\cite{Romans:1985tz}, which is a priori
unrelated to a gauging. However, let us reconsider the orthogonality relation
(\ref{eq:closure}), 
\begin{equation}
  \label{eq:closure-recall}
  \Theta_M{}^\alpha \, Z^{M}{}_{NP} ~=~0\,,
\end{equation}
which can be trivially satified by setting $\Theta_M{}^\alpha=0$.  In
view of the hierarchy~(\ref{sequence}), this deformation corresponds
to a sequence in which the lowest map $Y^{[0]}$ is absent such that
the hierarchy is not induced by the gauge interactions but starts at
the level of the 2-forms. It would be interesting to analyze the
general conditions under which such additional deformations can be
launched from higher-ranks in the hierarchy, e.g.\ under which
conditions the intertwining tensor $Z$ can contain representations
beyond those determined by the embedding tensor~(\ref{eq:def-Z}).

There is one other aspect that should be stressed. The gaugings are
controlled by the coupling constant $g$, and one may consider taking
the limit $g\to 0$. In that limit the covariant tensor hierarchy does
not reduce to a trivial abelian set of tensor gauge fields but also
reproduces non-trivial terms of order $g^0$. Consider as an example
the covariant field strength $\mathcal{H}_{\mu\nu\rho}{}^{MN}$,
defined in (\ref{eq:H-3-cov}), which contains Chern-Simons-like terms
that are not of order $g$. This feature, which may seem somewhat
surprising, was first noted in five-dimensional maximal supergravity,
where a Chern-Simons coupling is required by supersymmetry. However,
this Chern-Simons coupling is a special case of the Chern-Simons
coupling that is required by the gauge hierarchy \cite{deWit:2004nw}.
To put it differently, if supersymmetry would have excluded the
presence of a Chern-Simons coupling, then this theory could not have
been deformed by gauge interactions.

Finally, let us mention that for groups ${\rm G}$ other than the
series related to the maximal supergravities listed in
table~\ref{tab:vector-tensor-repr}, the tensor hierarchy that we have
exploited in this paper, may not run continuously all the way from
scalar fields to $d$-forms, but break off at some earlier stage.  This
happens e.g.\ for the groups ${\rm G}={\rm GL}(n)$ and ${\rm G}={\rm
  SO}(n,n)$ for which the hierarchy breaks off (upon imposing a mild
assumption regarding the representation constraints) after the vector
and the 2-form fields, respectively.  Accordingly, the associated
theories are not linked to specific space-time dimensions --- but
correspond to the $T^n$ torus reduction of pure gravity and bosonic
string theory, respectively, in an arbitrary dimension.  The
corresponding sequences~(\ref{sequence-symmetric}) will thus exhibit
an adequate gap in the middle, while the structure of forms with
$p\ge(d\!-\!3)$ remains the generic one that we have discussed in
section~\ref{sec:life-at-end}.  Another example in which the hierarchy
is degenerate concerns ten-dimensional IIB supergravity, which carries
only forms of even degree such that~(\ref{sequence-symmetric}) cannot
be established.

\newpage
\noindent
{\bf Acknowledgement}\\
\noindent
We are grateful to Hermann Nicolai for discussion. B.d.W. thanks the
Ecole Normale Sup\'erieure for hospitality extended to him during the
course of this work, which was supported by the Centre National de la
Recherche Scientific (CNRS). The work of H.S. is supported by the
Agence Nationale de la Recherche (ANR). This work is also partly
supported by EU contracts MRTN-CT-2004-005104 and MRTN-CT-2004-512194,
and by NWO grant 047017015. \bigskip


%
\end{document}